# Rock bottom, the world, the sky: Catrobat, an extremely large-scale and long-term visual coding project relying purely on smartphones

**Kirshan Kumar Luhana,** *kirshan.luhana@student.tugraz.at*
**Matthias Mueller,** *mueller@ist.tugraz.at*
**Christian Schindler,** *cschindler@ist.tugraz.at*
**Wolfgang Slany,** *wolfgang.slany@tugraz.at*
**Bernadette Spieler,** *bernadette.spieler@ist.tugraz.at*

Institute of Software Technology, Graz University of Technology, Austria

## Abstract

Most of the 700 million teenagers everywhere in the world already have their own smartphones, but comparatively few of them have access to PCs, laptops, OLPCs, Chromebooks, or tablets. The free open source non-profit project Catrobat allows users to create and publish their own apps using only their smartphones. Initiated in 2010, with first public versions of our free apps since 2014 and 47 releases of the main coding app as of July 2018, Catrobat currently has more than 700,000 users from 180 countries, is available in 50+ languages, and has been developed so far by almost 1,000 volunteers from around the world ("the world"). Catrobat is strongly inspired by Scratch and indeed allows to import most Scratch projects, thus giving access to more than 30 million projects on our users' phones as of July 2018. Our apps are very intuitive ("rock bottom"), have many accessibility settings, e.g., for kids with visual or cognitive impairments, and there are tons of constructionist tutorials and courses in many languages. We also have created a plethora of extensions, e.g., for various educational robots, including Lego Mindstorms and flying Parrot quadcopters ("the sky"), as well as for controlling arbitrary external devices through Arduino or Raspberry Pi boards, going up to the stratosphere and even beyond to interplanetary space ("the sky"). A TurtleStitch extension allowing to code one's own embroidery patterns for clothes is currently being developed. Catrobat among others intensely focuses on including female teenagers. While a dedicated version for schools is being developed, our apps are meant to be primarily used outside of class rooms, anywhere and in particular outdoors ("rock bottom", "the world"). Catrobat is discovered by our users through various app stores such as Google Play and via social media channels such as YouTube as well as via our presence on Code.org. Sharing, remixing, and collaboration is actively encouraged and supported. Catrobat has a very long term perspective in that it is independent of continuous funding and actively developed in a test-driven way by hundreds of pro-bono volunteers from around the world. Our aim is to grow by a factor of thousand and reach a billion users by 2030. We warmly welcome new contributors in every imaginable field and way with open arms. Please join us and contact me via [wolfgang@catrobat.org](wolfgang@catrobat.org) today!

## Introduction: Background, mission, and history

Knowledge in Computer Science (CS) is essential, and industries have increased their demand for professionals that have technical experience. The next generation of jobs will be characterized by new standards requiring employees with computational and problem solving skills in all areas, even if they are not actual technicians (Balanskat and Engelhardt, 2015). However, the number of young people, and women in particular, choosing to study and work in Information and Communication Technology (ICT) fields is decreasing dramatically (NCWIT, 2015; NCWIT, 2017; European Commission, 2016a). In the last decade, European technology employment has grown three times faster than all employment in total. The continuous improvements of technology and the numerous advancements in industrial processes made it possible to develop autonomous vehicles, robotics, 3D printing, genetic diagnostics, or Internet of Things (IoT) technologies. These technologies are already part of everyday life, and there is a corresponding growing worldwide need for qualified scientists, engineers, and technicians. For these reasons, society and governments have mandated that teenagers should acquire computing and coding skills (European Commission, 2016b), or even a new way of (critical) thinking and problem solving skills (Wing, 2006; Kahn, 2017; Tedre and Denning, 2016; Mannila *et al*., 2014). Presenting coding as a range of diverse skills which can be learned by adapting ideas from games is a generally applicable concept. Thus, a gamified and constructionist concept should hold teenagers' focus to actively participate by activating intrinsic and extrinsic motivators (Ryan and Deci, 2000). Games can be played everywhere, including on smartphones, tablets, and other digital devices. Moreover, the mobile game market continues to grow faster than other game industries, e.g., the number of game apps on Google Play grew by 28% in 2017 (Jingli, 2017; Takahashi, 2017).

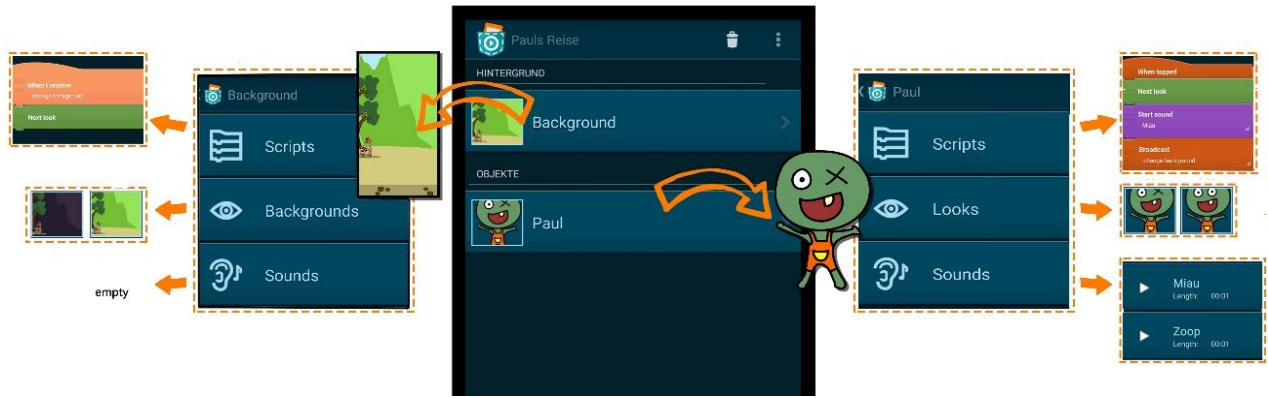

*Figure 1: Pocket Code's UI*

Catrobat's approach is inspired by Piaget's Constructivism theory 1948 (Piaget and Inhelder, 1967), starting with first computer programming courses at the MIT in 1962 (Greenberger, 1962), and refined with Papert's Constructionism concept in 1980 (Papert, 1985). Since then, different approaches were used to motivate kids for coding. With our free open source non-profit project Catrobat our goal is to provide computational thinking skills for everyone, especially teens from less developed areas where other computational devices such as PCs are almost non-existent.

Catrobat's apps and services have been immensely influenced by MIT's Scratch[1] project, and we consider Catrobat to be Scratch's little sister project for smartphones. Scratch itself has been strongly shaped by Papert's powerful ideas, and extends Papert's "Low Floor" but "High Ceiling" for the Logo programming language for kids (i.e., easy to start, but allowing to develop more complex projects as well) by adding "Wide Walls", emphasizing that Scratch supports a wide variety of projects as well as ways to learn and play, according to the needs and interests of its users (Resnick 2017). Going beyond Logo and Scratch's metaphor of the room, Catrobat's smartphone based approach allows to literally break down the walls of the room and move outdoors, thus inspiring Catrobat's mission statement, which is "Rock Bottom, the World, the Sky". "Rock Bottom" because on the one hand we aim at building upon and going beyond Scratch's focus on making the first experiences in coding as easy and satisfying as absolutely possible for our teenage user group, e.g., through a physics engine that is much easier and intuitive to use that similar concepts in other game making environments. On the other hand, because of our reliance on smartphones, it also is meant to evoke that our users can and, to a large percentage, do leave the "room" to code outside or create outdoor projects, in some cases literally "on the rock". Many of our users are indeed developing their projects while on the go, far from classrooms and their homes, and have developed apps that take advantage of the various sensors such as the cameras, GPS, compass, or acceleration sensors that are built into smartphones and that allow to create, e.g., augmented reality, geocaching, or dynamic outdoor sports games. Additionally, today's teenagers all over the globe have, to an already very large degree and also increasingly, their own smartphones readily available and permanently connected to the Internet, even in rural areas in Africa and other regions in the world where there may be no central power supply at home and kids charge their phones via solar panels in a central community facility of their village. Catrobat also has begun to become available in languages that are not supported by the phones makers themselves, such as Swahili, Gujarati, or Sindhi. Catrobat thus strives at reaching out to all corners of "the World" in an effective and efficient way that is indefinitely sustainable. The reliance on already existing smartphones, as well as the free open source character of Catrobat, which lets it thrive with little to no funding, also are significant aspects that allow Catrobat to scale up by avoiding the high costs and logistics that have hampered the success of similarly motivated projects in the past. Regarding the third part of our mission statement, Catrobat allows already since 2017 to program drones flying autonomously in "the Sky" (in particular, the popular Parrot Augmented Reality Drone 2.0), with a real-time video being transmitted to Pocket Code's screen, under full programming control by the user. While coding with Pocket Code has been done in airplanes, for the future we plan on going even higher. Because phones are small, lightweight, and portable, off-the-shelf Android phones have already been used to control balloons up to the stratosphere, and our extensions via Bluetooth and local WiFi connections to battery powered Arduino and Raspberry Pi allow to control any hardware of these flying computational systems, as long as the isolation keeps the harsh environment at bay and there's enough energy. On a further note, PTScientists' private enterprise Moon rover mission project[2], scheduled to lift off in 2019 and sponsored by, among others, Vodafone, Nokia Bell, Audi, and Red Bull, has several experiments on-board that rely on regular Android phones to lower the costs of otherwise extremely expensive "rocket science" hardware, with Vodafone and Nokia Bell sponsoring the installation of a 4G data network based on standard phone transmission technology from the rovers to the base station on the Moon and from there back to Earth. One of our dreams is that we will empower kids to use Pocket Code to design experiments and to program autonomous robots on the Moon, on Mars, and possibly even farther away. There is a thriving worldwide PhoneSat[3] community led by NASA that has launched a large number of nanosatellites based on 10x10x10cm cubesats using unmodified consumer-grade off-the-shelf Android smartphones and standard Arduino boards, which both would immediately work with Catrobat's apps. The sky has no limit, both in the concrete as well as in the metaphorical sense. Regarding the latter, our goal is to allow every kid to create complex, high resolution, and high performance apps using our tools, which they will be able to offer to other users, even commercially. Indeed, we allow our users to compile their apps into real Android apps, sign them with their own developer key, optionally add ads via their own AdMobs account, and sell them on Google Play for real money, thus directly empowering them to leave the limitations of the metaphorical "room" and bring their creations to the outside world.

---

[1] Scratch MIT: https://scratch.mit.edu/

On a historical note, one of the initial motivations for starting the Catrobat stems from Neal Stephenson's science fiction book "The Diamond Age: Or, A Young Lady's Illustrated Primer: a Propædeutic Enchiridion in which is told the tale of Princess Nell and her various friends, kin, associates, &c.". In the book, human tutors are hired anonymously on demand by a very affluent industrialist and aristocrat, to remotely educate a young girl, initially a toddler, living under gruesome conditions, by mistaking her, because of circumstances described in the book, for a princess for whom the primer was actually created. The illustrated primer, which is highly portable and has a voice- and touch sensitive interface that allows her to communicate with her tutors, accompanies Nell throughout her adolescence up to young adulthood, when she becomes a leading force and changes the fate of millions of the most underprivileged kids on Earth. One of the main story lines spanning a large part of the book is the acquisition of computational thinking skills by Nell through the Illustrated Primer. The book has won the Hugo and Locus science fiction awards, and was also cited by the developers of the One Note per Child Project as well as of Amazon's Kindle as a motivational inspiration. In 2017, Catrobat won the "Closing the Gender Gap" prize for a new subproject called "Remote Mentor", in which we have started to implement the necessary technical infrastructure and study the required social aspects to realize Stephenson's Illustrated Primer. We have partnered with sociology scientists focusing on gender aspects for the research part. In November 2017 we have begun to conduct initial remote mentoring experiments under real conditions, first with sponsoring from Google as a Google Code-in mentoring organization at the end of 2017, with several hundreds of teenagers from all over the world being remotely mentored by a pool of 38 Catrobat mentors over a period of seven weeks, followed by the "Remote Mentor" project itself, with initial funding from the Internet Foundation Austria until the end of 2018. Because of Stephenson's book, Catrobat has focused from its very beginnings to aim at empowering female teenagers in less affluent regions such as rural areas in India, Tanzania, or Brazil. Teens have reacted very positively to these first remote mentoring experiments, and we plan to eventually integrate the remote mentoring features into our tools and services so that mentors and mentees will be anonymously and automatically matched on demand, internationally in all languages, in a large scale, long term, and continuously further developed way.

---

[2] http://ptscientists.com/
[3] https://www.nasa.gov/phonesat/

This paper is organized as follows: First, we emphasize two trends that have emerged in the last decade (both were important in developing our app Pocket Code): block-based and visual coding, and an increased use of mobile devices among our main target group (teenagers from 13 to 19 years). Second, the focus lies on the Catrobat project and the educational app Pocket Code. Third, we describe our planned next steps and subprojects that are being developed, followed by a summary and conclusion.

# Computational Thinking Skills for All

Computational Thinking promotes the importance of coding and computer science activities, thus delivering concepts that are more applicable and highly essential to prepare teenagers for the future (Barnett *et al.*, 2017; Tetre et.al. 2016). After 2006, there was a rapid increase in the number of published articles about learning to code (Wu and Wang, 2012). The ongoing movement of promoting coding through visual programming languages has its origin at that time.

### Trend 1: Block-based and Visual Coding

In the last decade, a number of block-based visual programming tools, e.g., Scratch, have been introduced which should help teenagers to have an easier time when first practicing programming. These tools have all had very similar goals: they focus on younger learners, support novices in their first programming steps, they can be used in informal learning situations, and provide a visual/block-based programming language which allows teenagers to recognize blocks instead of recalling syntax (Tumlin, 2017). Unlike traditional programming languages, which require code statements and complex syntax rules, here graphical programming blocks are used that automatically snap together like Lego blocks when they make syntactical sense (Ford, 2009).

Another critical improvement of visual programming systems over classic text based programming languages is the fact that all elements of the programming environment and also the programming language itself, including the formula elements, are translated to the human language of the young users. Especially for human languages that are not written with the Latin alphabet, this is a huge advantage for users, as they are not used to think in English and very often have difficulties to even read Latin scripts. In case of developing countries, usually only a small percentage of the population understands English, in which most user interfaces are exclusively available, thus implicitly excluding a large part of the world's population. Localization of a software can revolutionize E-learning, resulting in more educated workforce and improved economy (Ghuman, 2017). Pocket Code supports localization and internationalization on the application level. The app's language and locale can be changed without changing the smartphone's interface language on the system level. Languages such as Sindhi and Pashto, which are yet to be supported by operating systems, can thus be seamlessly used by our users (Awwad, 2017). Catrobat shares this feature, which improves accessibility and inclusiveness to users from all regions of the world, e.g., with Scratch and Snap!, and this certainly contributes in a major way to the positive worldwide reception of these visual programming environments.

Thus, visual programming languages provide an easier start and a more engaging experience for teenagers. The ease of use, and simplicity of such programming environments enables young people to become game makers, and by the seamless translation of their user interfaces, to collaborate with other users on a worldwide scale.

### Trend 2: Smartphone Usage

With mobile games, more people can engage who were previously limited to use other platforms such as PCs or consoles. Further, children nowadays grow up with mobile devices and feel comfortable using them. Considering current prices, and the forecast of the user penetration of smartphones in Austria, France, Germany, and the United Kingdom from 2014 to 2021 (Statista Market Analytics, 2016), we can conclude that smartphones will be used significantly more by teenagers in the future than tablets, laptops, and desktop PCs. Smartphones and the use of apps are already a part of our culture and are changing the way in which many people, particularly teenagers, act in social situations. For most adolescents the smartphone performs several functions in their daily lives, and it contributes, e.g., to identity formation through self-presentation on the Internet. In addition, the smartphone is used a lot during spare time (most games are played in the evening (Verto Analytics, 2015)) or for just killing some time, e.g., when commuting. In addition, online games and mobile games play an important role in the daily lives of teenagers (Bevans, 2017). A recent study which examined American female players' experiences found that 65% of the Android users who play mobile games are women (Google and NewZoo, 2017).

Teenagers increasingly have mobile devices on their own, which enables them to creatively express themselves at any time and to use apps that bring their ideas and creations to life. With a more meaningful use of mobile devices, teenagers worldwide can acquire powerful knowledge that will make them into better problem solvers, thinkers, and learners.

## Catrobat and the Pocket Code App

The Free Open Source Software (FOSS) non-profit project Catrobat[4] was initiated 2010 in Austria at Graz University of Technology. The multidisciplinary team develops free educational apps for teenagers and programming novices. The aim is to introduce young people to the world of coding (Slany, 2014). With a playful approach, teenagers of all genders can be engaged, and game development can be promoted with a focus on design and creativity. A first public version of our free app was published in 2014, with 47 releases of the main coding app as of July 2018. Our app currently has more than 700,000 users in 180 countries, is natively available in 50+ languages (including several languages not directly supported by the underlying operating system), and has been developed so far by almost 1,000 volunteers from around the world.

These volunteers are implementing software, designing educational resources, translating the app, or provide other services that help to advance the project. The contributors work together in a cooperative way, having the chance to engage in a field they like and create something within a community that follows the same shared vision. Besides attracting students, educators, and other interested contributors from all over the world, Catrobat also benefited from being part of Google's Summer of Code and Code-In initiatives. These initiatives promote open source worldwide and motivate teenagers and students to get involved in projects such as ours. Our openness towards international contributors helps us to represent different cultures, bring in various viewpoints, and generate new ideas how the project can further develop. Catrobat's project management as well as development is done in an agile way, allowing our contributors to adopt new technologies, respond to user feedback, and embrace upcoming ideas quickly. An example where a feature request issued by users was implemented is our web based automatic APK (Android Package) generation. It is currently being extended for users to be able to sign their apps and add AdMobs based ads, so that they can publish their Catrobat projects on Google Play and other app stores and earn money with them. Another example is of technical nature and affects our deployment workflow, which was redesigned and fully automated to enable us to deploy to Google Play including the localized screenshots and app descriptions in 47 languages (not all our languages are supported on the Google Play store) with only two mouse clicks. All this helps to provide a motivating user experience for our young target group, support them in their learning process, and foster collaboration.

---

[4] https://www.catrobat.org

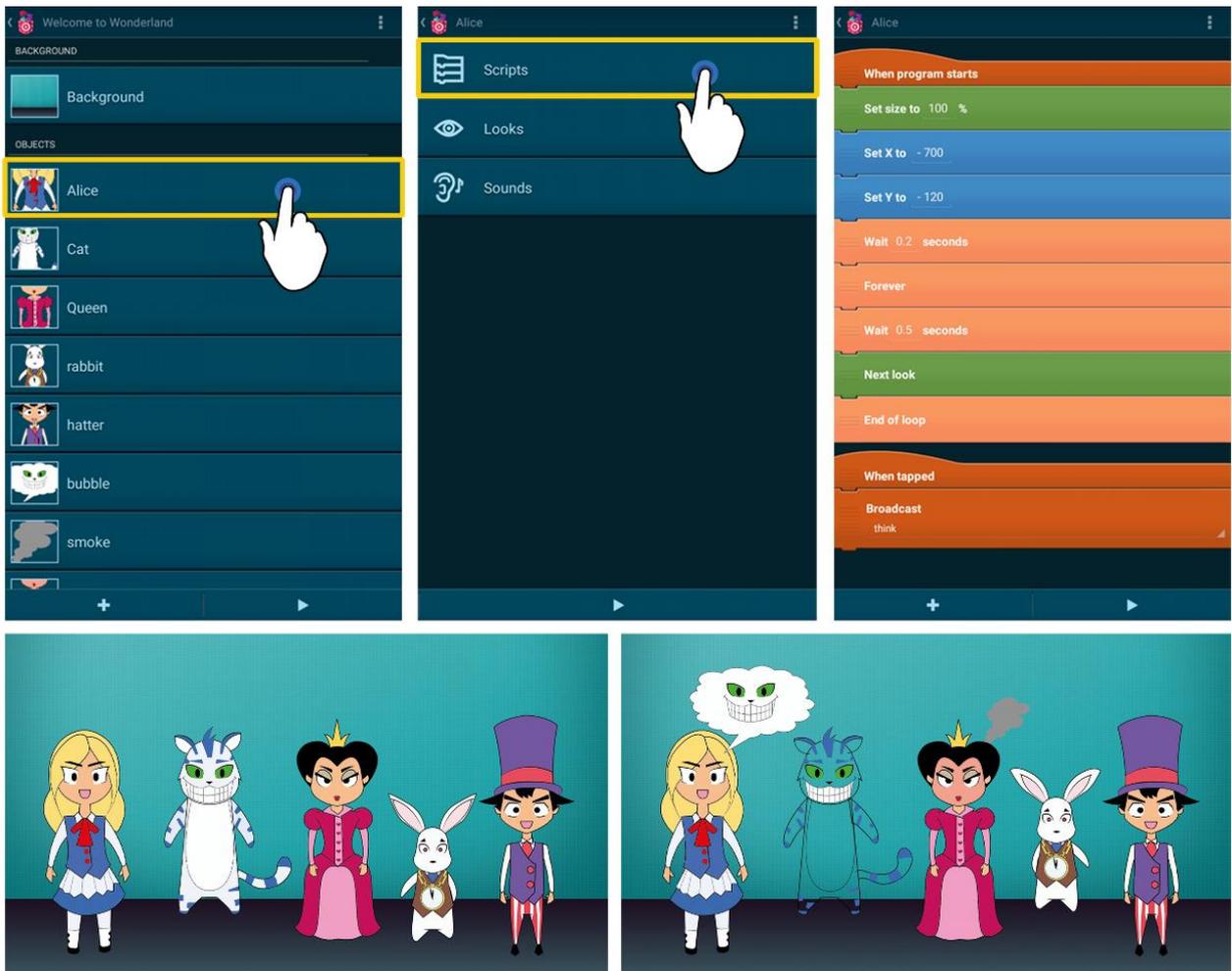

Figure 2: Pocket Code Alice themed program

## Pocket Code: Creating your own Games

The app Pocket Code[5] is an Android-based visual programming language environment that allows the creation of games, stories, animations, and many types of other apps directly on phones or tablets, thereby teaching fundamental programming skills. This app consists of a visual Integrated

---

[5] https://catrob.at/pc

Development Environment (IDE) and a programming language interpreter for the visual Catrobat programming language. The IDE automatically translates the underlying code parsed by the XML file into visual brick elements and vice versa. With the use of simple graphic blocks, users can create their own game, colorful animations, or extensive stories directly on the mobile phone without prior knowledge.

The drag and drop interface provides a variety of bricks that can be joined together to develop fully fledged programs. The app is freely available for Android on Google's Play Store and soon will be available on Apple's iTunes Store for iPhones. Figure 2 shows Pocket Codes' UI and an example project with "Alice in Wonderland" characters.

## Pocket Code: the Mobile Integrated Coding Environment

Projects in Pocket Code follow a similar syntax to the one used in Scratch and are created by snapping together command bricks. They are arranged in scripts that can run in parallel, thereby allowing concurrent execution. To communicate between objects, to trigger execution of scripts, or scripts beyond objects, broadcast messages are used. By means of this mechanism, sequential or parallel execution of scripts is possible, either within the same object or over object boundaries. In addition to the basic control structures, Pocket Code offers event triggering building blocks for event-driven programming. Familiar concepts, such as variables, lists, or Boolean logic, are included as well.

In addition to Scratch, Pocket Code has a 2D physics engine which enables the user to define certain physical features of objects and the stage (collision detection, velocity, gravity, mass, a bounce factor, and friction) to create from simple up to complex simulations of the real world. An example that showcases both the physics engine as well as the use of inclination sensors is a simulated wooden maze through which a metal ball needs to be navigated from a starting position towards the winning end position, all while avoiding a number of holes on the floor of the maze, by tilting one's phone, as if it were a real, physical wooden maze. The wooden walls initially execute a physics brick named "Set motion type" with the pull-down option "others bounce off it", and the ball executes the same "Set motion type" brick with the option "bouncing with gravity". The movement of the ball by tilting the phone is realized by a "Set gravity for all objects to X: -3 x inclination_x  Y: -3 x inclination_y steps/second$^2$" brick that is executed in a "Forever" loop. Voilà, that's all that is needed. To increase the realism of the simulated maze, there is an additional "When physical collision with anything" brick followed by a "Vibrate for 0.02 seconds". Even more, the ball's metallic reflection sheen is oriented always in the same direction using the magnetic compass sensor of the phone, thus giving the impression that the light always comes from the same side. It would be easy to also influence the gravity vector using the acceleration sensors built into the phones, which would make the ball not only react to tilt, but also to shaking and quick moves in any direction. The rest of the scripts handles the animations when the ball "falls" into one of the holes, and also shows the amount of holes that have successfully been avoided so far in the top left corner. Note that all objects that execute one of the two variants of the "Set motion type" brick mentioned above, automatically have their convex hull computed for the physics collisions, based on the visible parts of their current look. No other game engine to our knowledge does this automatically: in other gaming frameworks, the bouncing box is either quadratic, or has to be manually specified by a professional developer. Catrobat's use of convex contours makes it extremely intuitive and simple for the user to create complex games using the physics engine. Note that arbitrary 2D forms going beyond the convex hull, e.g., patterns of the form "U" or "8", are much more complex to handle automatically, and it is probably easier for users to handle special cases on an individual basis, e.g., by forcing several objects to move in synchronicity. In contrast, for all practical purposes it is impossible to realize a physically correct collision and bouncing of objects from each other for arbitrary shapes in other visual programming languages such as Scratch. This aligns with the "rock bottom" metaphor of Catrobat corresponding to the "low floor" metaphor of Logo and Scratch, making complex games easy and intuitive to realize, with a very low entry threshold, with physics being a concept everyone is intimately familiar with. At the same time, the maze project also has a resolution of 1920 x 1080 of which every pixel is fully used, giving it a very high-resolution and also highly realistic look, on par with professionally created game apps. This exemplifies Catrobat's "the sky" metaphor corresponding to the "high ceiling" of Logo and Scratch, since with Pocket Code, there literally is no upper limit in realism and performance of the created games: a project's resolution is only limited by the capabilities of the phone on which it is created, so, e.g., a project with a 3840 x 2160 pixels resolution can be created using a Sony Xperia XZ Premium smartphone. Figure 3 shows the stage as well as some of the scripts mentioned above. The project can be found under the name "Tilt maze 1.0" on Catrobat's sharing site from within Pocket Code.

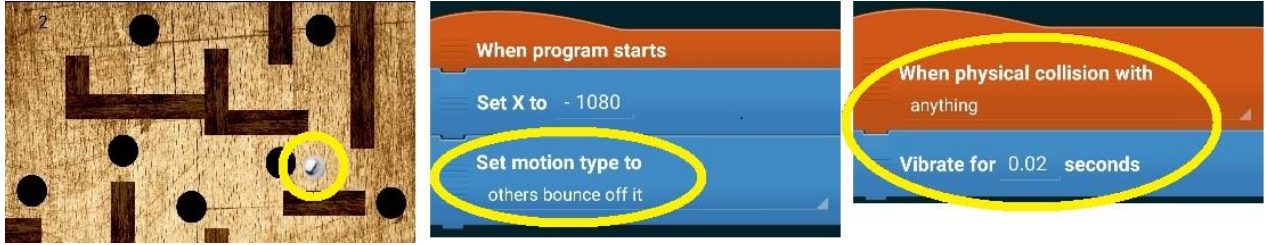

*Figure 3: Partial screenshots from "Tilt maze 1.0" that relies on Pocket Code's physics engine, where the physical behaviour of objects is set through its "motion type", e.g., "others bounce off it" for the wooden walls of the maze (middle script), and a short vibration when the ball touches a wall (script on the right). The movement of the ball by tilting the phone is realized by a "Set gravity for all objects to X: -3 x inclination_x Y: -3 x inclination_y steps/second$^2$" brick that is executed in a "Forever" loop (not shown here). Direct link to the "Tilt maze 1.0" project on the web version of Catrobat's sharing site https://catrob.at/TiltMaze*

With Pocket Code's intuitive merge functionality, the new parts of two projects can be seamlessly merged together into one larger project – parts of the two initial projects that exist in both are not duplicated. This makes programming cooperatively much easier. Modern smartphones are equipped with a large number of sensors, although most mobile games only use few or none of them (Kafai and Vasudevan, 2015). Within Pocket Code, users can create games using the device's sensors, such as inclination, acceleration, loudness, face detection, GPS location, or the compass direction, which makes user input easy and engaging. With Pocket Code it is also possible to connect via Bluetooth to Lego® Mindstorms robots or Arduino™ boards. The following extensions are available: Lego Mindstorms NXT/EV3, Parrot AR.Drone 2.0 and Parrot Jumping Sumo Drone, Arduino, Raspberry Pi (via WiFi), NFC tags, Phiro robots[6], and Chromecast. These kinds of computational construction kits make creating programmable hardware accessible to even novice designers and combines coding and crafting with a rich context for engaging teenagers (Kafai and Vasudevan, 2015). In the context of robots, being able to program a smartphone makes much more sense, as the smartphone can be mounted on the robots, thus allowing to give it a face, a voice and other sounds, and additional sensors such as acceleration, inclination, magnetic field, GPS, voice recognition, computer vision. Also, since only a smartphone is needed with Catrobat, the programming can be done on the spot, outside, e.g., when using one's land-based robot or flying drone outdoors. With Catrobat no laptop or PC is necessary, thus, coding can take place anytime and anywhere, and in particular can be widely made available even in less affluent communities around the world. In addition, Catrobat released a Scratch Converter to allow the conversion of existing Scratch projects to the Catrobat language directly within the app, so there are, in fact, now more than 30 million projects available for remixing and inspiration to our users.

The Pocket Code interface consists of several very distinct areas: First, the app itself with a main menu and the collection of downloaded or developed-by-oneself projects, second a community sharing[7] platform, which is integrated into the app as a web-view, and which serves as a learning, sharing, remixing, cooperation, and publishing place, third the "stage" where projects get executed on the phone, and additionally a sophisticated graphical editing program that allows to draw and edit the looks of all actors, objects, and backgrounds of one's projects. This community website provides an online platform for users to download and upload programs, share them with other users, search for programs, and to provide feedback, e.g., write a comment to a project or rate a project. In addition, tutorials and starter projects are provided. In the community website's project overview, users can execute the project directly in desktop browsers (HTML5 web player), download the project to the Pocket Code app, or download the project as a standalone Android app. In addition, the tool automatically creates statistics from Pocket Code projects and provides an online code overview. The project details page is illustrated in Figure 4. Figure 5 illustrates the options within the main menu.

---

[6] https://catrob.at/Phiro
[7] https://share.catrob.at

*Figure 4: Pocket Code web share: project details page with code statistic and code view*

If the user starts first with a new and empty program, it initially only contains one empty background object. With the "+" sign users are able to add objects, looks, or sounds (depending on which activity he or she is in). The background object itself can be assigned to several backgrounds, which can be exchanged during runtime. The background can also have its own scripts. Every project can consist of multiple objects and at least one background (which is a special kind of object). Every object can hold a.) scripts that define the behavior of the object, b.) looks which can be changed and used, e.g., for object animation, and c.) sounds to make the object play music, other sounds, or recorded speech. Scripts can control the looks and sounds. Looks can be drawn and edited with Pocket Paint. Pocket Paint[8] is a second app of Catrobat available on Google Play, which allows users to create their own objects with a pencil or different shapes (note that we currently work on integrating this second app completely in Pocket Code in order to simplify the installation for our users). Distinctive features of Pocket Paint include the ability to use transparency, to zoom in up to pixel level, to change the dimensions of the looks, and to use layers, the latter being particularly interesting to create consecutive looks from an animation series. In addition, users can add looks with their camera, from their phone's memory, or use Catrobat's Media Library with a collection of predefined graphics. To add a new sound the user can either record a sound directly in Pocket Code, add a sound from the Catrobat Media Library, or add a sound from the phone's memory. This workflow is illustrated in Figure 6.

---

[8] https://catrob.at/PPoGP

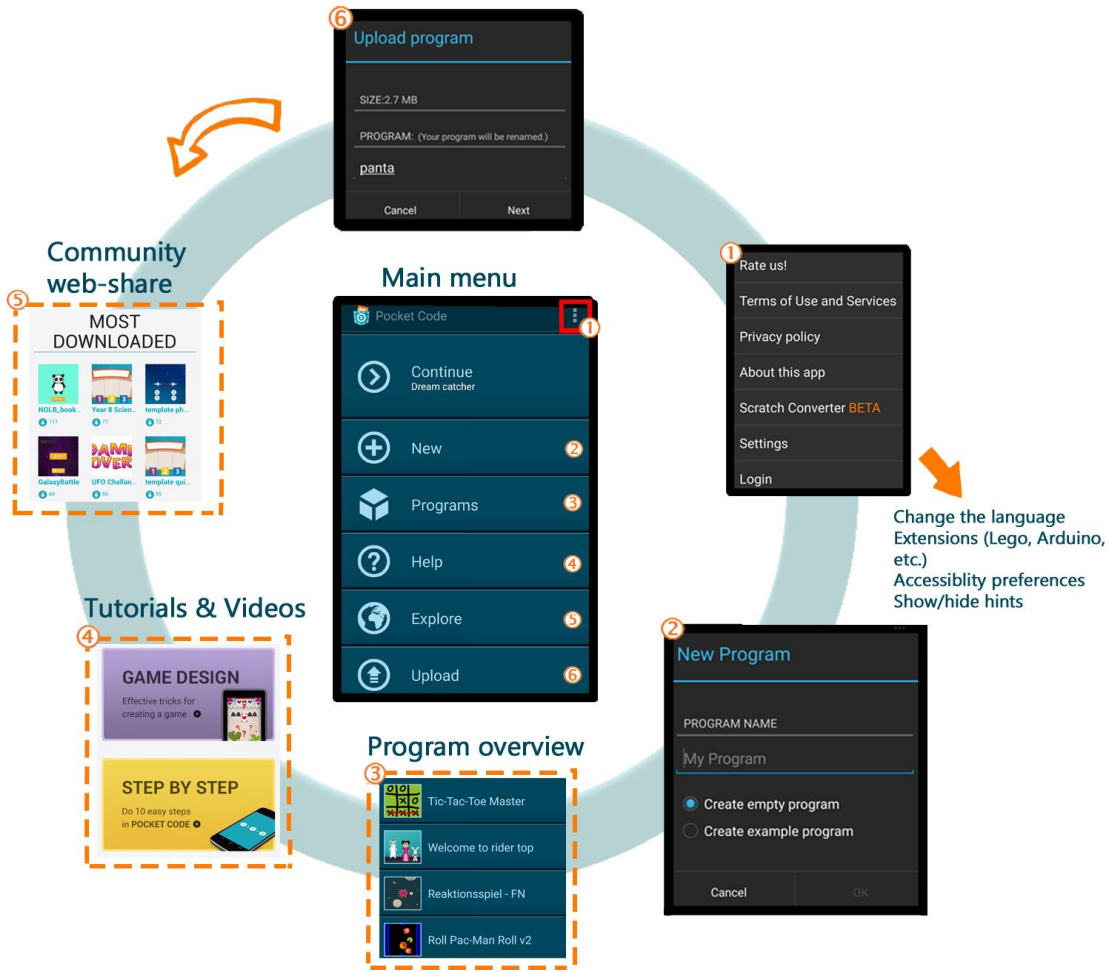

*Figure 5: Pocket Code main menu: within the settings menu you can find, e.g., the accessibility preferences or the Scratch Converter; 2) create a new project by starting with an example game or with an empty game; 3) project overview: tap on one to execute or modify it; 4) find help: videos, tutorials, step-by-step tutorials, education page for teachers and students or google groups forum; 5) download and play games from other users; 6) upload your game to the sharing platform.*

A script is a collection of code blocks that contain the logic of programming and define the operations of the object. Thus, it is possible to move the object and access its properties and change them. For adding scripts there are seven different brick categories (see Figure 7): a.) The *Event* category in dark orange that contains hat-bricks or broadcast bricks. Hat bricks are special kinds of bricks that, depending on certain circumstances such as a tap on an object, start the attached script; b.) The *Control* category in orange contains if-then-else bricks, loop-bricks to control the flow of the script, bricks to switch between scenes, clone bricks, etc; c.) The *Motion* category in blue color contains bricks to manipulate the object's position, orientation, or movements; d.) The *Sound* category in purple contains bricks to start and stop sounds, manipulate the volume, or accept spoken input; e.) the *Looks* category in green contains bricks to change the graphical appearance of the object, e.g., set/change size, brightness, transparency or hide/show the object as well as set a certain look to animate the object, to show speech and think bubbles, or to ask for written user input; f.) The category *Pen* in dark green holds bricks for drawing lines (a pen that follows the object) and the option to leave stamped marks of the object on the background, g.) The *Data* category in red contains bricks to manipulate variables and lists, e.g., to set/change variables, maintain lists, add/insert/replace items, and show variable content on the stage. This color scheme makes is possible to understand scripts more easily through of the bricks' color which supports readability. By activating extensions in the settings menu, additional categories appear for Lego (yellow), Drone (brown), Arduino, the Phiro robot (both in cyan), etc.

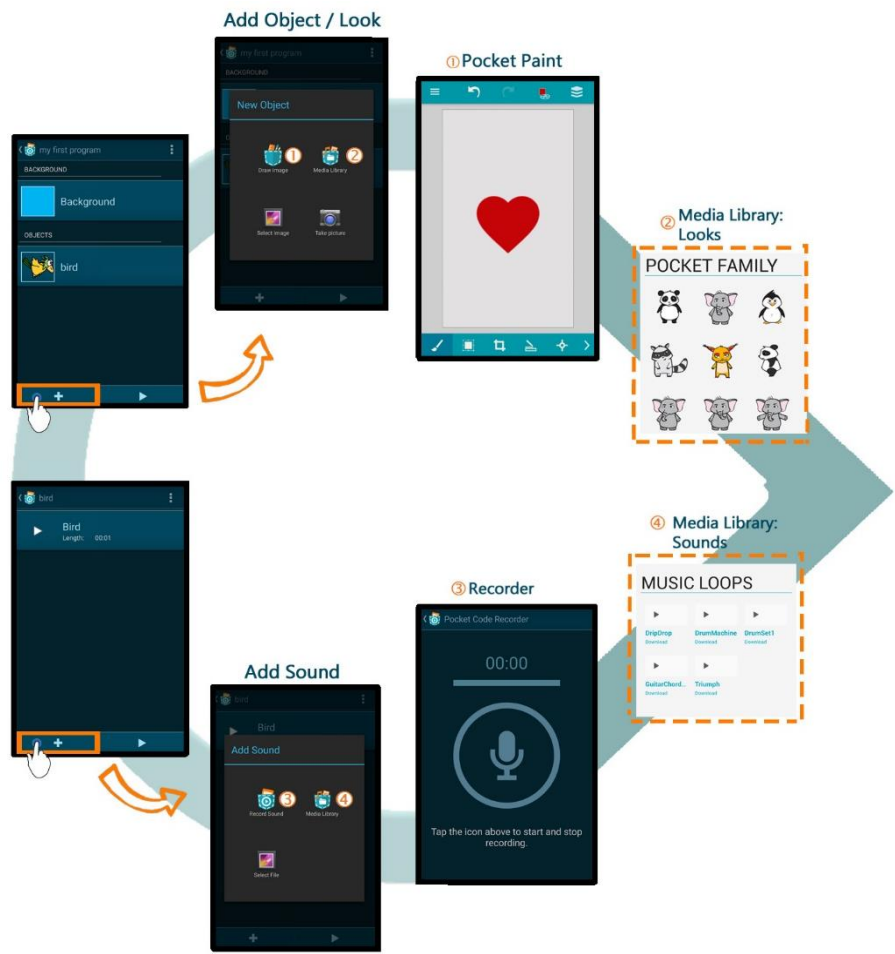

*Figure 6: Pocket Code's UI: add a look/object or add a sound with the "+" sign.*

In contrast to Scratch and Snap!, Pocket Code does not use bricks for formulas. Instead, there is a **formula editor** that looks like a calculator and allows the creation and execution of mathematical and logical formulas that can be used in bricks. In (Harzl et al. 2013) we compared of blocks-based (like in Scratch and Snap!), text-based (like with traditional programming languages such as Python), and hybrid ways (such as Pocket Code) to enter and edit formulas, and showed that teenagers can create and debug complex formulas using a hybrid editing mode faster and with less errors than with the blocks-based as well as purely text based approaches. All advantages of block-based interaction modes, such as easy discoverability of features, translation into many human languages, avoidance of syntax errors, and immediate feedback about current values (including dynamic sensor values), are not only present in Pocket Code's formula editor, but are additionally enhanced by the familiarity of how formulas are written, changed, and read by users in other contexts, as well as by the familiar interface of an electronic pocket calculator (app). Pocket Code's formula editor is shown in Figure 8. It consists of an input field to show and compose the formula, a keyboard, and a compute button to display the current result. On the keyboard, five categories for various values, functions, and operators are available. a) *Object*: a collection of values of the current object, e.g., values for the X- and Y-coordinate, or the current speed, b) *Functions*, such as sin or cos, a random number generator, or list and string functions, c) *Logic* is used to compare values or to combine logical expressions, d) in *Device* there is information that the smartphone or tablet records, e.g., inclination, loudness, or GPS data, and e) *Data* stores created variables and lists and shows their last value.

With a tap on the play button the program starts. The objects are shown on **the stage** and the scripts are executed. To stop or to pause the program, the user has to tap on the back button of the phone. A stage menu appears which can be seen in Figure 9. The stage is organized in a logical coordinate system with an X- and Y-axis, which allows an exact positioning of the objects. This axis can be displayed in the stage menu (see Figure 9c).

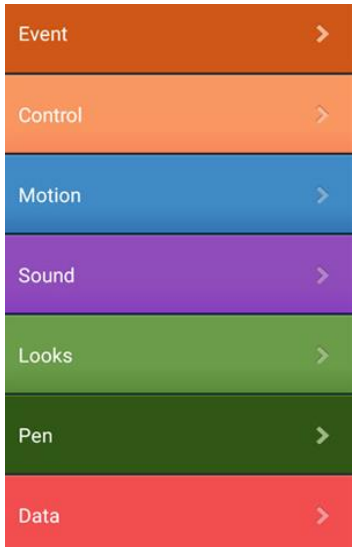

*Figure 7: Script categories: choose bricks from the seven basic available categories.*

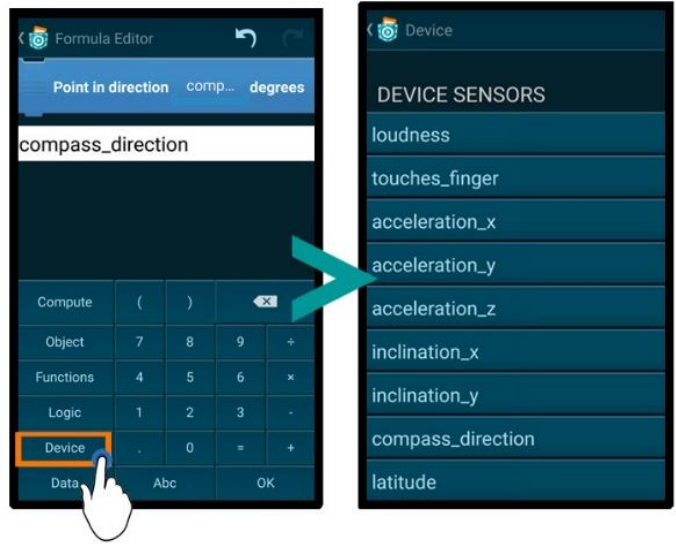

*Figure 8: Formula editor: the value for the direction can be defined as a constant or, e.g., a sensor can be chosen by tapping on "Device".*

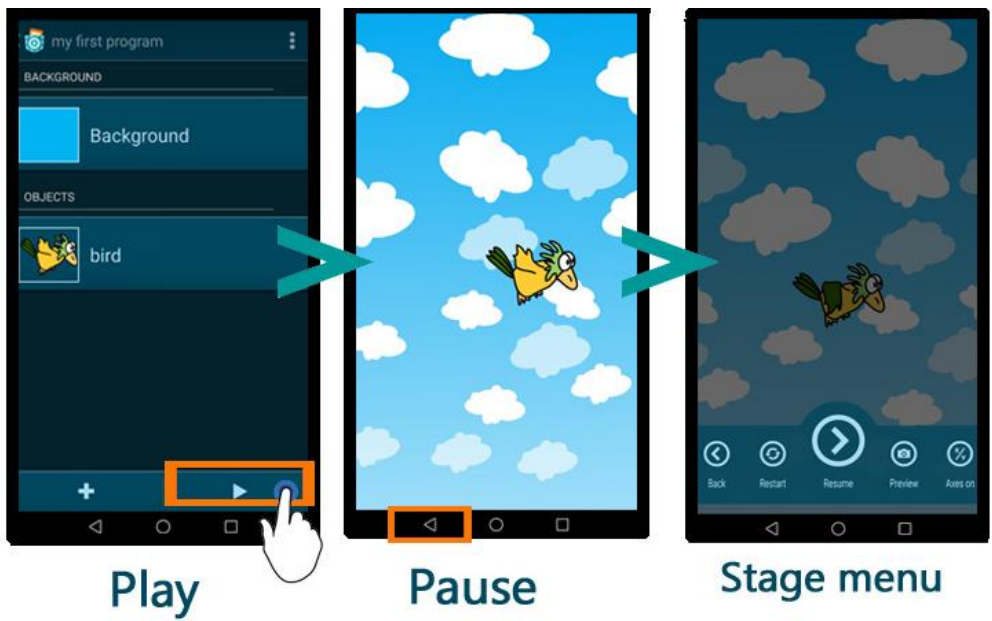

*Figure 9: Stage; a.) tap the play button to start the program, b.) tap the back button of the phone to pause the game, c.) in the stage menu the user has five options: 1. tap back again to stop the game and switch*

back to editing of project, 2. restart the game, 3. resume the game, 4. add a new preview picture to the project (this will be shown, e.g., on the sharing platform), and 5. display the x/y axes on the device screen.

# Projects and Further Work

This new and forward-thinking approach to code on mobile devices received national and international recognition. The Catrobat project has won a number of awards, including 2016 two Lovie Awards ex aequo with Red Bull and Doctors without Borders, evaluating the best European digital projects in London, and the Reimagine Education Award for innovative educational projects at the Wharton Business School of Pennsylvania[9]. Additionally, in March 2017, Catrobat won the "Platinum Award" in Best Mobile App Awards Best Educational App category and 2016 the "Internet for Refugees"[10] award for a Right-to-Left language implementation of Pocket Code, which supports several RTL languages, e.g., Arabic or Farsi, and particularly focuses on refugees and teenagers in crisis or development areas. A new project was started in January 2018 which promotes remote mentoring by connecting female role models with female programming beginners. This idea was awarded with the "Closing the Gender Gap" prize of the Austrian NetIdee in November 2017. During the European H2020 project "No One Left Behind" (NOLB), the team developed a special flavoured school version of the app with the name "Create@School". This version compromises, e.g., the gathering of analytics data for visualization, the integration of accessibility preferences, and the development of pre-coded templates. Currently a new flavored version customized for female teenagers is in development. This version with the name "Luna&Cat" promotes special content for girls, like featured and user-contributed programs, media assets, and tutorial videos.

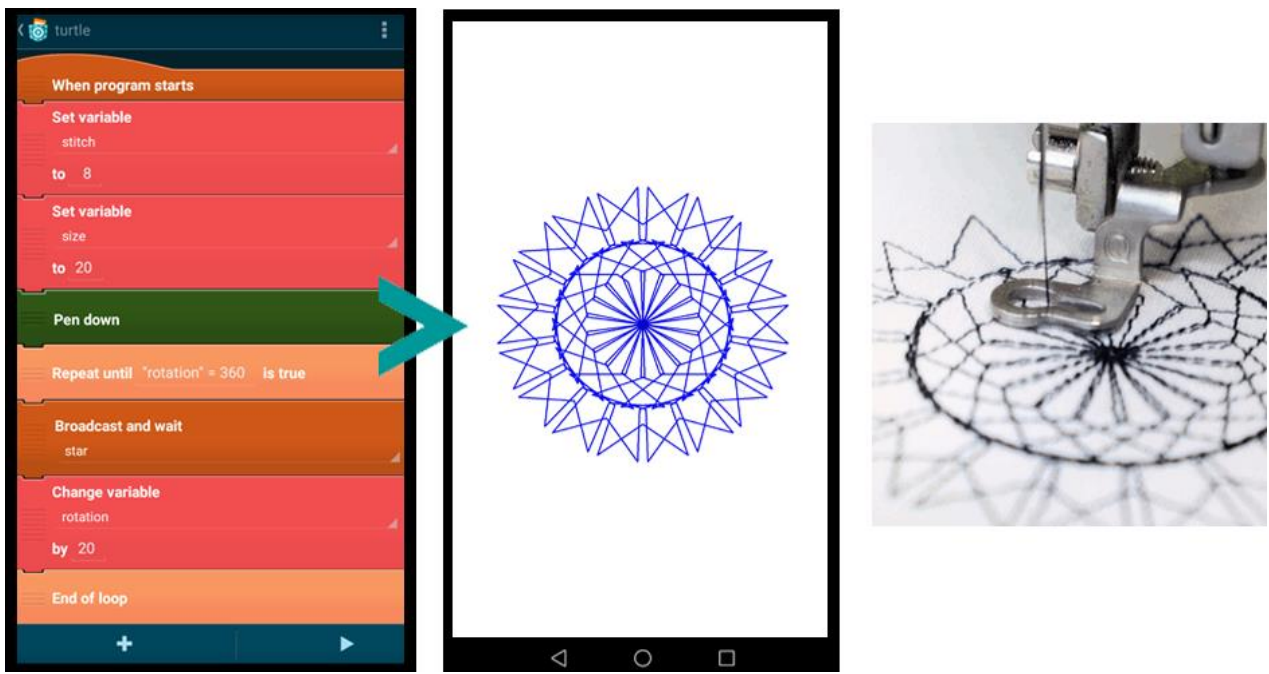

*Figure 10: "Stitched" patterns in Pocket Code. Picture on the right with kind permission from Andrea Mayr-Stalder, www.TurtleStitch.org project.*

One new feature which is currently under development is an extension to program embroidery machines. Once available, self-created patterns and designs can be stitched on t-shirts, pants, or even bags or shoes. With Pocket Code, the embroidery machines will be programmable, similar

---

[9] http://www.reimagine-education.com/awards/reimagine-education-2016-honours-list/
[10] http://www.tugraz.at/en/tu-graz/services/news-stories/tu-graz-news/singleview/ article/preis-internet-for-refugees-fuer-programmier-app-der-tu-graz

to the existing TurtleStitch[11] project, which realizes this concept on a PC (while with Pocket Code only a smartphone is needed). As a result, teenagers have something they can be proud of, something they can wear, and they can show to others. This feature has proven to be especially engaging for female teenagers and shows them new ways of expressing themselves creatively through coding. Figure 10 shows an example of an embroidery pattern made with Pocket Code.

Another new beta feature allows registered users to sign and release the Catrobat project as apps on Google Play. Users optionally also add mobile ads to earn money. In developing countries, mobile technologies are playing an important role in developing economies (Alderete, 2017). This is because mobile phones and the mobile internet require considerably fewer financial resources in comparison to a traditional desktops and laptops (Stork, 2013). Due to lack of alternate employment opportunities in developing countries, the low cost of investment is a critical enabling factor for new entrepreneurs (Alderete, 2017). The economic advantage of releasing an app on Google Play or integrating AdMob within apps can motivate many people to learn programming and solve issues digitally. This is especially beneficial for under-privileged user groups who have access to limited resources like computers and continuously available electricity. They can benefit by sharing their creativity with others and serve a global market with minimal resources such as a low-cost smartphone and mobile internet, which are increasingly available everywhere. Figure 11 shows an example of apps with AdMob integration.

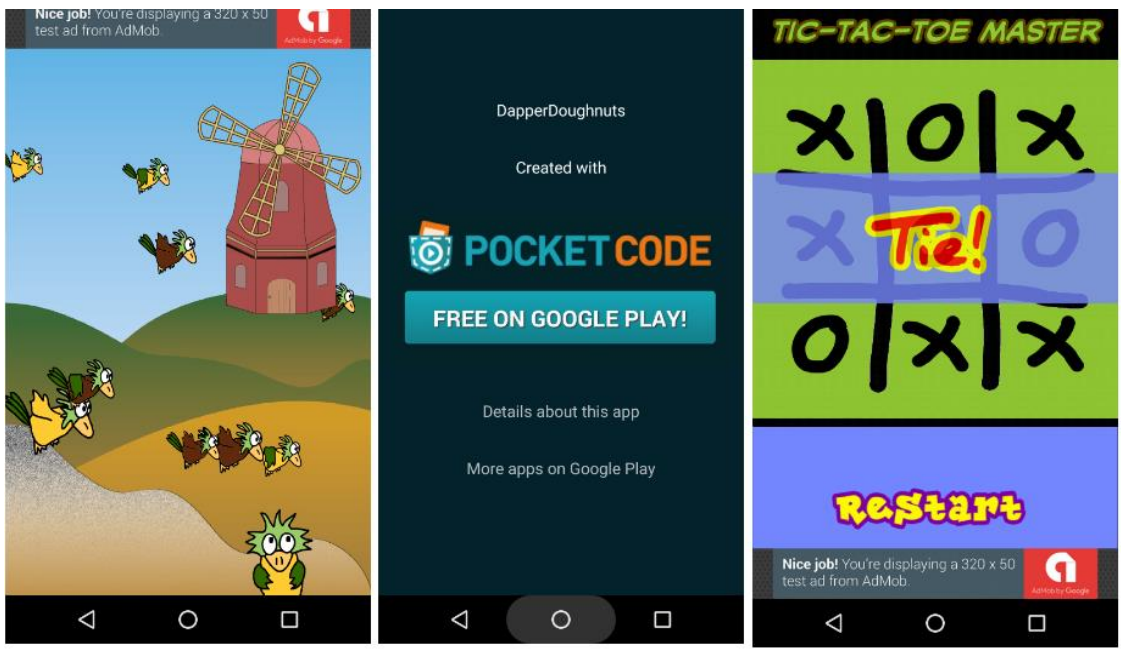

*Figure 11: Android apps with AdMob banner advertisement created with Pocket Code.*

## Conclusion and Discussion

The aim of this paper was to provide an overview about the Catrobat project and the Pocket Code app as of July 2018. Pocket Code provides an easy way to start coding. It is not intended to allow the development of standard applications, but to promote understanding of the logic behind coding and foster computational thinking skills, thus following a constructionist approach in learning by doing and the creation of sharable artefacts. Creative and artistic talents can be recognized and learning can occur in a user-centered, project-based setting with the use of new media. Users of Pocket Code are mostly teenagers who can learn from each other and share their ideas to create new games and other apps together. The community sharing platform allows users to give and receive feedback, support, and assistance from others around the world, thus allowing our users to stand on the shoulders of their peers and learn from each other. They can try out new ideas and realize the projects they define for themselves, aided and inspired by likeminded others in a user-friendly and social environment. Catrobat fosters diversity and learning in a worldwide community. Our goal is to empower teenager all over the world to realize their potential and express themselves creatively with today's and any future technology.

---

[11] http://www.turtlestitch.org/